\documentclass[prb,showpacs,twocolumn]{revtex4}
\usepackage{amsmath}
\usepackage{amssymb}
\usepackage{bm}
\usepackage{epsfig}
\usepackage{graphicx}
\usepackage{color}
\usepackage{natbib}
\usepackage{hyperref}
\usepackage{epsf}
\usepackage{amssymb}

\usepackage{color}

\begin{document}

%\pnum{}

\title{Resonant photonic crystals and quasicrystals based on \\ highly doped
quantum-well structures}

\author{M.~M.~Voronov, N.~S.~Averkiev, M.~M.~Glazov}

\affiliation{Ioffe Physical-Technical Institute RAS, 26 Polytekhnicheskaya, 194021 St.-Petersburg, Russia}

\begin{abstract}
{A theory of light propagation through one-dimensional photonic
crystals and deterministic aperiodic structures, including quasicrystals,
based on doped quantum-well structures has been developed.
The resonant Bragg condition, leading to the superradiant regime
and formation of the widest optical reflection spectrum, has been
formulated. The expressions for band gap edges for light waves in the
Bragg structures have been obtained. The reflection and absorption spectra of such systems are calculated.
The optical properties of the doped multiple-quantum-well structure are compared with the properties of
undoped ones.}
\end{abstract}

\maketitle
\section*{Introduction}
The control of light propagation and light-matter interaction on the nanoscale are among the rapidly developing directions of the solid state physics. The systems allowing the constructive interference of light waves in a certain frequency range are of special interest. The pathway to achieve the emission control is to implement ordered multiple quantum well (MQW) structures consisting of thin semiconductor quantum wells in dielectric matrix. The examples of such systems are one-dimensional (1D) photonic crystals with periodic arrangement of MQWs [1] and quasicrystals [2] which are non-periodic but yet possessing long-range order that leads to coherent Bragg diffraction of propagating waves [3].

The considerable interest is attracted by the resonant structures where  the Bragg diffraction takes place at frequencies corresponding to the excitonic resonances of the system. This is the case of undoped structures or systems with moderate doping. In a highly doped QW, bound exciton states vanish owing to both the screening of the Coulomb potential and the
state-filling, and an interband absorption demonstrates the Mahan singularity [4]. 

Here we focus on the MQW systems consisting of $N$ highly doped QWs with their centers positioned at points $z=z_m$ $(m=1...N)$, which form periodic or quasiperiodic sequence. We calculate reflection and absorption spectra of these systems governed by the Mahan singularity for both periodic and quasi-crystalline systems and study the transition between the superradiant and photonic crystalline regimes of the light propagation in MQWs.  We demonstrate that the spectra of doped MQW systems are strongly asymmetric with respect to the absorption edge frequency as compared with the systems with exciton resonances.

\section{Periodic MQW structure}

We start with expressions for the light reflection coefficient from a single doped QW, which takes the form~[5]
\begin{equation}\label{1}
r_{<} =
\frac{\rm{i}{\Gamma}^{\alpha}}{(\omega_0-\omega)^{\alpha}-
\rm{i}{\Gamma}^{\alpha}}\:,\:\:\:{\rm if}\:\:\omega \leq
\omega_0
\end{equation}
\begin{equation}\label{1}
r_{>} = \frac{\rm{i}{{\Gamma}}^{\alpha}
e^{\rm{i}\pi\alpha}}{(\omega-\omega_0)^{\alpha}-
\rm{i}{\Gamma}^{\alpha}
e^{\rm{i}\pi\alpha}}\:,\:\:\:{\rm if}\:\:\omega \geq \omega_0
\end{equation}
Here $\omega$ is the light frequency, $\omega_0 = (E_g + E_F)/\hbar$ is the
absorption edge frequency,
with $E_g$ being the band gap energy and $E_F$ being the Fermi energy, the parameter $\alpha$ ($0<\alpha<1$) is a function of the electron density, determining the strength of the Mahan singularity. Constant $\Gamma = [C\Gamma_0/\sin(\pi\alpha)]^{1/\alpha} E_F/\hbar$ characterizes the light-matter coupling strength, where
$\Gamma_0$ is the exciton radiative damping in the undoped quantum well,  $C=\pi a^2m/(4\hbar)$, and $a$ is the two dimensional Bohr radius. A specifics of the system under study is the interaction of the photon mode with the continuum of electron-hole pair excitations in quantum wells.

There are two limiting regimes of the light propagation in 1D
resonant photonic crystals: (i) superradiant regime, which can be
realized for small numbers of wells $N$ leading to constructive
Bragg interference and (ii) photonic crystal regime, which is
realized for large enough values of $N$, so that forbidden band gap
is already essentially formed~[1]. The dispersion equation for
polaritons propagating through one-dimensional photonic crystal can
be reduced to the following form~[1]:
\begin{equation}\label{1}
\cos Qd = D_1 + \frac{\mathrm{i} r}{1+r}D_2 \:,
\end{equation}
where $Q=Q(\omega)$ is wave vector of polariton, $D_1 = \cos qd$ and $D_2 = \sin qd$ for the structures with negligible mismatch of dielectric constants, with $d$ being the structure period and $q(\omega)=\omega n/c$, $n$ is the refractive index. Hereafter the normal incidence of light is assumed.
An analysis of the dispersion Eq.~(3) with the reflection
coefficient in form of Eqs.~(1), (2) shows that if $q(\omega_0)d =
\pi m$, $m=1,2,...$ which we call the resonant Bragg condition
(RBC), then there exists a wide band gap in energy spectrum for
light waves in infinite photonic crystal. In case of $m=1$ we get
the widest band gap whose edges correspond to the ``polaritonic"
wave vector $Q=\pi/d$ and on frequency scale lie between the value
$\omega_{-} = \omega_0
-(2\omega_0\Gamma^{\alpha}/\pi)^{1/(1+\alpha)}$ and $\omega_0$.
Unlike the undoped system at $\omega>\omega_0$ the reflectivity of the system
decreases because the strong absorption caused by the formation of free
electron-hole pairs arises.

Now we turn to the superradiant regime realized at $N\ll 1/{\rm
Re}|(Qd)-\pi|, 1/|{\rm Im}(Qd)|$ and present the expressions for
intensity reflection coefficient, $R_<$ (for $\omega<\omega_0$) and
$R_>$ (for $\omega>\omega_0$)
\begin{equation}
R_< = \frac{N^2{\Gamma}^{2\alpha}}{(\omega_0 -
\omega)^{2\alpha}+N^2{\Gamma}^{2\alpha}} \;,
\end{equation}
\begin{equation}
R_> = \frac{N^2{\Gamma}^{2\alpha}}{(\omega -
\omega_0)^{2\alpha}+N^2{\Gamma}^{2\alpha}+2(\omega -
\omega_0)^{\alpha}N{\Gamma}^{\alpha}\sin(\pi\alpha)} \:.
\end{equation}
Figure~1 shows typical reflection spectra $R_N$ from doped MQW structures with
different numbers $N$ of wells, calculated under resonant Bragg
condition. According to  Eq.~(4) the halfwidth of left part of the
spectra up to the value $N\approx 64$ is increasing proportionally
to $N$. For larger $N$ the increase of the spectral width is
superlinear due to the band gap formation. The asymmetry of the
spectra with respect to the resonant frequency $\omega_0$ is a
specific feature of doped MQW structures.
In our calculations we disregard the inhomogeneous broadening of the
electron states. The absorption spectra $A_N=1-R_N-T_N$, where
$T_N$ is intensity transmission coefficient for the same systems, are
shown in Fig.~2.
The strong asymmetry of the absorption coefficient of doped MQWs is seen as
compared with the undoped system with excitonic resonances.

\begin{figure}
\leavevmode\epsfxsize=2.7in\centering{\epsfbox{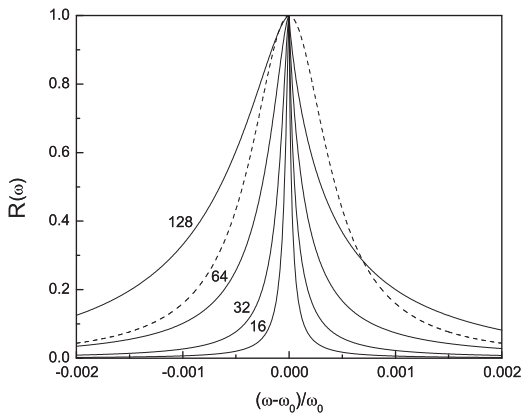}}
\caption[]{Light reflection spectra from doped MQW-structures ($\alpha =
0.8$) with $N = 16, 32, 64 $ and $128$ wells compared to the
spectrum from exciton MQW-structure (dashed line). Calculated for
$\hbar \Gamma_0 = 10$ $\mu$eV, $\hbar \omega_0 = 1.5$ eV under RBC.}
\end{figure}

\begin{figure}[h]
\leavevmode\epsfxsize=2.7in\centering{\epsfbox{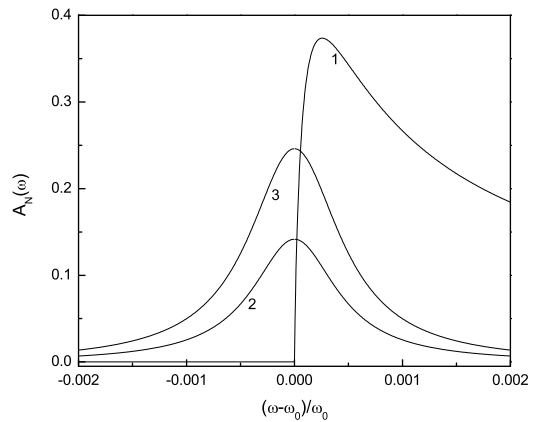}}
\caption[]{The absorption spectra (1) for doped MQW-structure
($\alpha=0.8$) compared to absorption spectra (2), (3) for exciton
MQW's calculated under RBC, when $\hbar \Gamma_0 = 10$ $\mu$eV and
$\hbar \omega_0 = 1.5$ eV. The spectra 2 and 3 are calculated for
the non-radiative exciton damping $\gamma = 5\Gamma_0$ and $\gamma =
10\Gamma_0$, respectively.}
\end{figure}

\section{Quasicrystalline MQW structure}

Now we turn to the quasicrystalline MQW structures. As an example we consider
Fibonacci sequence of the quantum wells arranged as follows.
We denote by $A$ and $B$ the segments consisting of a quantum well and adjacent
barrier of width $a$ and $b$, respectively. A quasicrystal can be
constructed by using so-called substitution rules for segments $A$ and $B$,
which in our case (Fibonacci) read as $A\to AB$, $B\to A$. Thus, we have
a sequence of wells and barriers, beginning with $ABAABABA...$.

\begin{figure}
\leavevmode\epsfxsize=2.7in\centering{\epsfbox{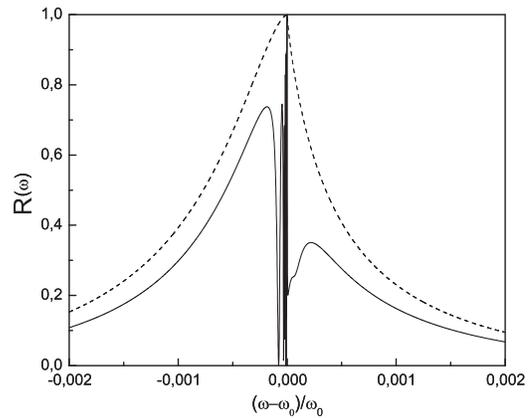}}
\caption[]{Reflection spectrum for doped MQW-quasicrystalline structure calculated for $\rho=\sqrt{2}$ under RBC.
%with  (h,h')=(1,0).
Dashed line shows spectrum for periodic MQW-stucture
with the same $\bar{d}$. In both cases $\alpha =0.8$, $N = 144$, $\hbar
\Gamma_0 = 10$ $\mu$eV and $\hbar \omega_0 = 1.5$ eV.}
\end{figure}

Here we extend the theory of light propagation through 1D resonant photonic
quasicrystals developed in [2] for the case of highly doped quantum wells with
Mahan singularity. Figure 3 shows the reflection spectrum for a noncanonical
Fibonacci QW-structure with ratio of the barrier widths $\rho\equiv
b/a=\sqrt{2}$ and the average period $\bar d$ satisfying resonant Bragg
condition $\omega_0 n \bar{d}/c = \pi$. There are two peaks in the spectrum,
on the left and right side from frequency $\omega_0$, which can be
described with high accuracy in two-wave approximation, as in
Ref.~[2]. The left peak lies in a photonic band gap,
which plays a role of usual band gap for energy states in a periodic
structure. The right peak is distinctly lower than the left one
because of higher absorption, c.f. Fig.~2. The peaks in the vicinity
of $\omega=\omega_0$ form a self-similar structure and are smeared
out with allowance for the inhomogeneous broadening. Dashed line in
Fig.~3 shows the reflection spectrum calculated for periodic MQW
structure with the same average period and other parameters as for
the quasicrystalline QW-structure.

In conclusion, we have shown that highly doped MQW structures can give strong effects of Bragg reflection and superradiance. Contrary to the systems with excitonic resonance the spectra of doped MQWs are essentially asymmetric allowing finer tuning of optical properties.

\acknowledgments We acknowledge the support by RFBR, President grant for young scientists,
the ``Dynasty'' Foundation -- ICFPM, Linkage Grant of IB of BMBF at DLR
and Russian Ministry of Education and Sciences.


\begin{thebibliography}{3}
\itemsep-2pt

\bibitem{Iv}
E.L.~Ivchenko et al., {\em Phys. Rev. B} {\bf 70}, 195106 (2004).

\bibitem{Po}
A.N.~Poddubny et al., {\em Phys. Rev. B} {\bf 80}, 115314 (2009).

\bibitem{Ya}
C. Janot, Quasicrystals: A Primer (Clarendon, Oxford, UK, 1994)

\bibitem{Ma}
G.D. Mahan, {\em Phys. Rev.} {\bf 153}, 882 (1967); {\bf 163}, 612 (1967).

\bibitem{Av}
N.S.~Averkiev et al., {\em Phys. Rev. B} {\bf 76}, 045320 (2007).

\end{thebibliography}
\end{document}